\documentclass[aps,prl,preprint]{revtex4}

\usepackage{graphicx}
\usepackage{amsmath}

\begin{document}


\title{Single-charge occupation in ambipolar quantum dots}


\author{A. J. Sousa de Almeida}
\email[]{a.j.sousadealmeida@utwente.nl}
\affiliation{NanoElectronics Group, MESA+ Institute for Nanotechnology, University of Twente, the Netherlands}
\author{A. M\'{a}rquez Seco}
\affiliation{NanoElectronics Group, MESA+ Institute for Nanotechnology, University of Twente, the Netherlands}
\author{T. van den Berg}
\affiliation{NanoElectronics Group, MESA+ Institute for Nanotechnology, University of Twente, the Netherlands}
\author{B. van de Ven}
\affiliation{NanoElectronics Group, MESA+ Institute for Nanotechnology, University of Twente, the Netherlands}
\author{F. Bruijnes}
\affiliation{NanoElectronics Group, MESA+ Institute for Nanotechnology, University of Twente, the Netherlands}
\author{S. V. Amitonov}
\affiliation{NanoElectronics Group, MESA+ Institute for Nanotechnology, University of Twente, the Netherlands}
\author{F. A. Zwanenburg}
\affiliation{NanoElectronics Group, MESA+ Institute for Nanotechnology, University of Twente, the Netherlands}


\date{\today}

\begin{abstract}
We demonstrate single-charge occupation of ambipolar quantum dots in silicon via charge sensing.
We have fabricated ambipolar quantum dot (QD) devices in a silicon metal-oxide-semiconductor heterostructure comprising a single-electron transistor next to a single-hole transistor. Both QDs can be tuned to simultaneously sense charge transitions of the other. We further detect the few-electron and few-hole regimes in the QDs of our ambipolar device by active charge sensing.

\end{abstract}


\maketitle

\section{Introduction}



The spin state of a single electron or single hole confined to a semiconductor quantum dot (QD) provides a promising system for quantum computation~\cite{Loss-PRA-1998}.
From the several contenders for coherent and scalable spin qubits, spins in silicon QDs have proven particularly appealing~\cite{Veldhorst-NatN-2014}. Silicon is the standard material for complementary metal-oxide-semiconductor technology, which promises to ease the implementation and scalability of solid-state qubits towards industrial applications~\cite{Zwanenburg-RMP-2013,Veldhorst-NatN-2014,Maurand-NatC-2016}.
Furthermore, natural silicon consists predominantly of zero nuclear magnetic moment isotopes, suppressing spin dephasing via hyperfine interaction~\cite{Elzerman-Nat-2004,Petta-Sci-2005,Koppens-Nat-2006,Camenzind-NatC-2018}. This enables long spin coherence times in comparison to III$-$V semiconductors. Electrons in silicon also experience weak spin-orbit interaction so that their spins are largely immune to charge noise~\cite{Kuhlmann-NatP-2013}. 
These properties have prompted extensive research on electron spins in Si QDs for quantum computing. Si QDs are highly sensitive electrometers, enabling charge transfer signals of a QD down to single-electron occupation~\cite{Yang-AIP-2011} and high-fidelity single-shot spin readout via spin-to-charge conversion~\cite{Elzerman-Nat-2004}.
The manipulation of electron spins in Si QDs is commonly achieved via spin resonance techniques~\cite{Pla-Nat-2012,Kawakami-NatN-2014,Wu-PNAS-2014,Hao-NatC-2014,Veldhorst-NatN-2014}. However, these methods require the presence of static magnetic field gradients from micromagnets for electric dipole spin resonance~\cite{Kawakami-NatN-2014,Wu-PNAS-2014} or microwaves for electron spin resonance~\cite{Pla-Nat-2012,Hao-NatC-2014,Veldhorst-NatN-2014}, which are difficult to apply to individual spins in multiple qubit devices and thus may compromise the device scalability.

Holes in Si QDs have attracted significant attention for spin qubits due to the possibility of performing fast, highly coherent qubit operations~\cite{Szumniak-PRL-2012,Kloeffel-PRB-2013,Brunner-Sci-2009,Greve-NatP-2011}.
In contrast to electrons, hole spins in silicon have inherently strong spin-orbit coupling due to the $p$-wave symmetry of their Bloch wavefunction. This enables spin control using local electric fields applied by gate electrodes~\cite{Pribiag-NatN-2013,Golovach-PRB-2006,Flindt-PRL-2006,Nowack-Sci-2007} and potentially fast spin manipulation times~\cite{Maurand-NatC-2016,Watzinger-NatC-2018}. Further, holes in silicon experience small hyperfine coupling to nuclear spins, which has been predicted to yield 10 to 100 times enhancement of T$_2$ over electron spins in Si~\cite{Bulaev-PRB-2005}.
Despite these promising properties, hole spins in Si QDs have remained mostly unstudied. Single-hole occupation has been reported in silicon nanowires~\cite{Zhong-NL-2005} and only very recently in planar silicon QDs~\cite{Liles-SR-2018}.

While it remains unclear whether the electron spin or the hole spin in silicon is most suitable as a qubit, ambipolar devices allow the confinement and manipulation of both spin types in the same crystalline environment and in a single device~\cite{Betz-APL-2014,Mueller-NL-2015,Kuhlmann-APL-2018,Spruijtenburg-SR-2016}. This enables the direct comparison of electron and hole spin properties and benchmarking which is more suitable for spin qubits. 
Ambipolar device operation has been previously demonstrated in field-effect transistors integrating both $n$- and $p$-type reservoirs on the same device~\cite{Betz-APL-2014,Mueller-NL-2015,Spruijtenburg-SR-2016}, or a metallic nickel silicide compatible with standard complementary metal-oxide-semiconductor fabrication~\cite{Kuhlmann-APL-2018}.
These studies reported operation of electron and hole quantum dots in Coulomb blockade regime~\cite{Betz-APL-2014,Mueller-NL-2015,Kuhlmann-APL-2018}, as well as the improvement of device performance via passivation of charge defects by annealing in a H$_2$ atmosphere~\cite{Spruijtenburg-SR-2016}.
Ambipolar devices have so far been studied in the many-charge regime via direct transport measurements, due to the difficulty in depleting their QDs to the single-spin regimes.
The reason for this limitation is that, with decreasing number of confined spins, the tunnel barriers defining the QD become extremely opaque and the transport signal drops abruptly~\cite{Liles-SR-2018, Betz-APL-2014}. An alternative method for studying electrical transport in QDs is to use one QD as a sensor to charge displacements in another nearby~\cite{Field-PRL-1993}. This method has been used to detect the single occupation of electron~\cite{Yang-AIP-2011} and hole~\cite{Liles-SR-2018} QDs and thus may well be suitable to detect the few-charge regime in ambipolar devices.

Here, we report the implementation of ambipolar charge sensing in Si QDs. Our device comprises a single-electron transistor (SET) and a single-hole transistor (SHT) in a planar Si structure. The electron and hole QDs can sense charge displacements in the other, and our device can be tuned so both QDs are sensing each other simultaneously.
We further implement a charge sensing method with a feedback control loop to operate the charge sensor at constant current and sensitivity~\cite{Yang-AIP-2011}.
Using this method we demonstrate the few-electron and few-hole occupation of both the SET and SHT.

\section{Experimental details}

\begin{figure}
\includegraphics[width=9cm]{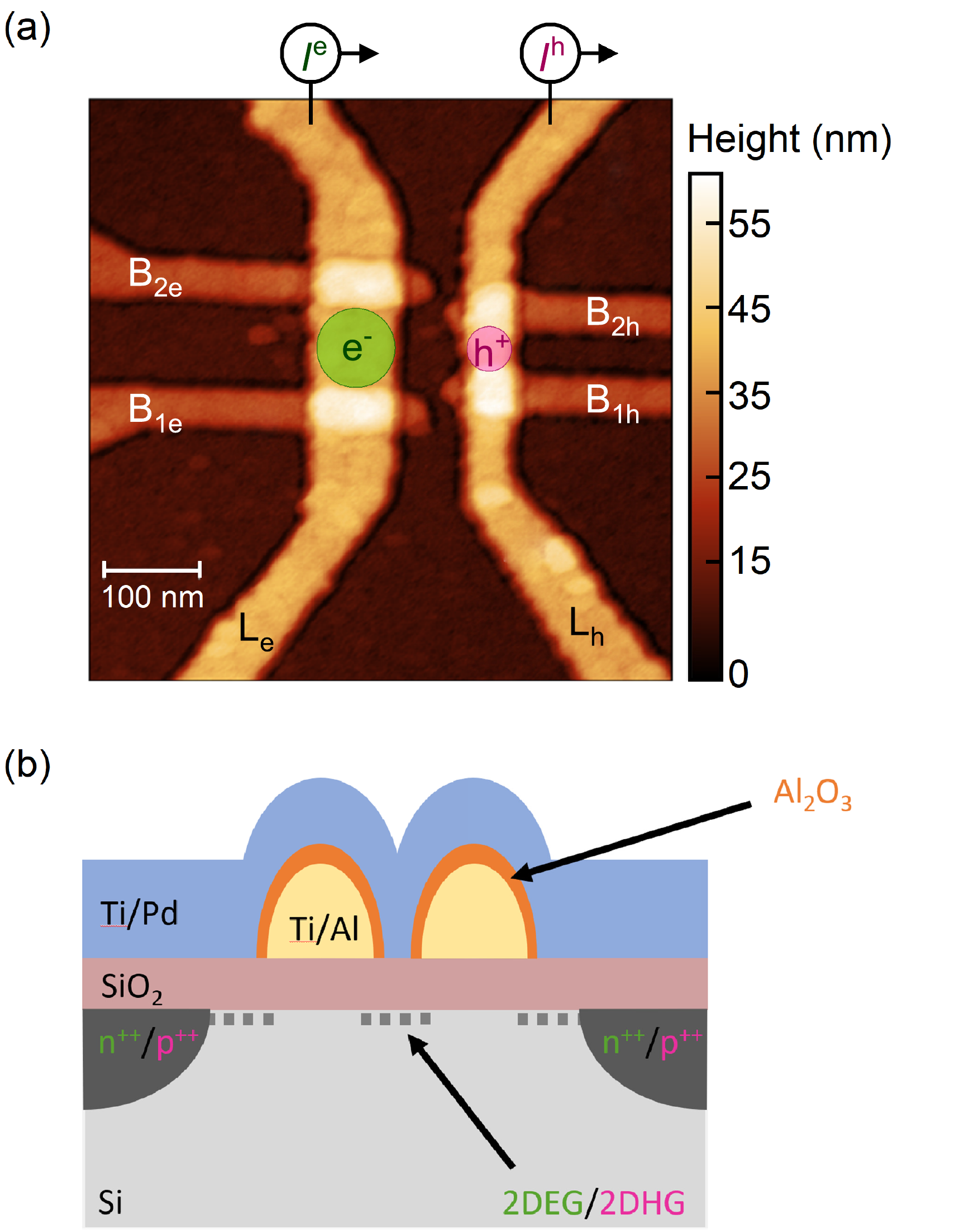}
\caption{Ambipolar quantum dot device. (a) Atomic force micrograph of the device, showing the SET (left) and SHT (right) regions. Each region comprises three gate electrodes: two barrier gates which create tunnel barriers to the QD, and a lead gate which applies the voltage needed to form a two dimensional electron (hole) gas at the Si/SiO$_2$ interface. (b) Schematic cross-section of the device. Dark grey represent the electron and hole reservoirs. \label{fig:fig1}}
\end{figure}

We have developed an ambipolar device in a silicon-based MOSFET-type heterostructure consisting of a single-electron transistor (SET) capacitively coupled to a single-hole transistor (SHT). These two regions are defined electrostatically by means of gate electrodes which control charge accumulation at the Si/SiO$_2$ interface. 
Figure~\ref{fig:fig1} shows an atomic-force microscopy image and a schematic cross-section of our ambipolar device, which was made with a combination of optical and electron-beam lithography. We use an intrinsic Si(100) wafer ($\rho \geq 10$~k$\Omega$) as a substrate. Source and drain regions used as electron (hole) reservoirs are implanted with phosphorus (boron) dopant atoms. A layer of 7.5~nm thermally grown silicon oxide is used as an insulating barrier between the substrate and the gate electrodes. Two layers of gate electrodes are patterned using electron-beam lithography. The first of these layers comprises Ti/Al (0.5 nm/35 nm) barrier gates with a typical width of 35~nm and a separation between barrier gates of $\sim$100~nm for the SET and $\sim$40~nm for the SHT. After deposition this layer is thermally oxidized to form a 
layer of Al$_2$O$_3$. 
The second layer comprises Ti/Pd (1/60 nm) lead gates of the SET and of the SHT, which are used to provide a conducting path from each source to the corresponding drain. 
Following the creation of the gate layers, the sample is annealed in hydrogen at 400$^o$C to passivate defects at the Si/SiO$_2$ interface~\cite{Spruijtenburg-Nanotechnology-2018}.
Transport measurements are performed in a dilution refrigerator with a base temperature of 10~mK and an effective electron temperature of $\approx$25~mK~\cite{Mueller-RevSciInstr-2013}. All voltages are given with respect to ground.

\section{Results and discussion}

\begin{figure*}
\centering
\includegraphics[width=16cm]{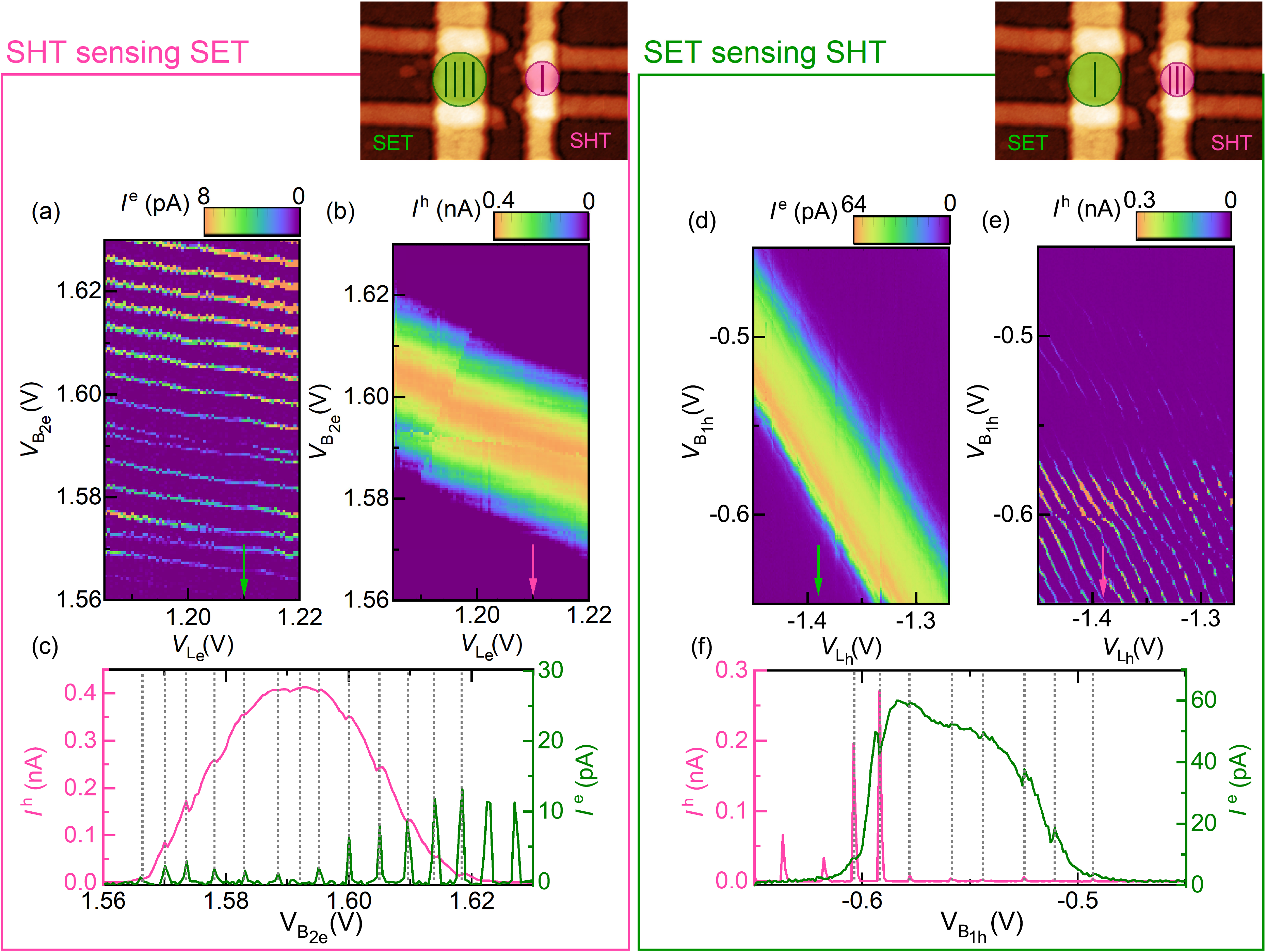}%
\caption{Ambipolar charge sensing. (a)$-$(c) Charge sensing using the SHT as sensor. Source-drain current versus $V_{\text{B}_{2\text{e}}}$ and $V_{\text{L}_e}$ through (a) the SET (labeled $I^e$) and (b) the SHT (labeled $I^h$). Data acquired at $V_{\text{B}_{1e}}=1.12$~V, $V_{\text{L}_{h}}=-1.45$~V, $V_{\text{B}_{1h}}=-0.70$~V, and $V_{\text{B}_{2h}}=-0.53$~V. (c) Line traces of $I^e$ and $I^h$ at the values of $V_{\text{L}_e}$ indicated by the arrows in (a) and (b), respectively. 
(d)$-$(f) Charge sensing using the SET as sensor. Source-drain current versus $V_{\text{B}_{1\text{h}}}$ and $V_{\text{L}_h}$ through (d) the SET (labeled $I^e$) and (e) the SHT (labeled $I^h$). Data acquired at $V_{\text{B}_{2h}}=-0.55$~V, $V_{\text{L}_{e}}=1.83$~V, $V_{\text{B}_{1e}}=0.98$~V, and $V_{\text{B}_{2e}}=1.16$~V. (f) Line traces of $I^e$ and $I^h$ at the values of $V_{\text{L}_h}$ indicated by the arrows in (d) and (e). The schemes at the top-right of each panel represent the alignment of the SHT and SET levels for each charge sensing regime. SET and SHT source-drain voltages were fixed at 0.5~mV.  \label{fig:fig2}} 
\end{figure*}

To investigate ambipolar charge sensing in our device, we
first study the linear transport regime by simultaneously measuring the source-drain current through the SET $I^e$ and through the SHT $I^h$.
In the left panel of Fig.~\ref{fig:fig2} we use the SHT to sense charge transitions in the SET.
Fig.~\ref{fig:fig2}(a) shows the charge stability diagram of the SET. A pattern of regularly spaced Coulomb oscillations highly coupled to $V_{\text{BR}_e}$ indicates the formation of an electron QD in the SET.
We tune the SHT to a single Coulomb oscillation [see Fig.~\ref{fig:fig2}(b)] and then sweep two gate voltages controlling the SET, $V_{\text{B}_{2e}}$ and $V_{\text{L}_e}$. 
The height of the hole peak is modulated by a regular pattern of abrupt upsets.
In the plots of linecuts of $I^{\text{e}}$ and $I^{\text{h}}$ at $V_{\text{L}_e}=1.21$~V voltage [see Fig.~\ref{fig:fig2}(c)], the Coulomb oscillations in the SET match the locations of the abrupt ridges
in the hole peak.
Thus, we infer that the SHT is sensing single electron transitions in the SET~\cite{Yang-AIP-2011}. %

In the right panel of Fig.~\ref{fig:fig2} we use the SET as a charge sensor for the SHT. As in the reciprocal regime described above, we tune the SET to a single Coulomb oscillation and sweep two gate voltages controlling the SHT, $V_{\text{B}_{1h}}$ and $V_{\text{L}_h}$, while measuring $I^{\text{e}}$ and $I^{\text{h}}$ simultaneously. The charge stability diagram of the SHT in Fig.~\ref{fig:fig2}(e) shows regularly spaced Coulomb oscillations. 
The spacing between electron peaks of the SET in Fig.~\ref{fig:fig2} is smaller than between hole peaks. This indicates that the size of the hole QD is smaller than the electron QD, thus corresponding to the lithographic dimensions of our ambipolar device. 
The charge stability diagram of the SET shown in Fig.~\ref{fig:fig2}(d) displays a single Coulomb oscillation with intensity modulated by the pattern of hole charge transitions observed in $I^{h}$, as becomes clear from the linecuts of $I^{\text{e}}$ and $I^{\text{h}}$ at $V_{\text{L}_h}=-1.41$~V in Fig.~\ref{fig:fig2}(f). 
In this figure, we also see that hole peaks below the noise level of our direct-transport measurements are still detected as upsets in $I^{\text{{e}}}$. 
We have demonstrated in Fig.~\ref{fig:fig2} the possibility of ambipolar charge sensing using either the SET or the SHT to sense single charge transitions in the other region of the device.

%

\begin{figure}
\centering
\includegraphics[width=13cm]{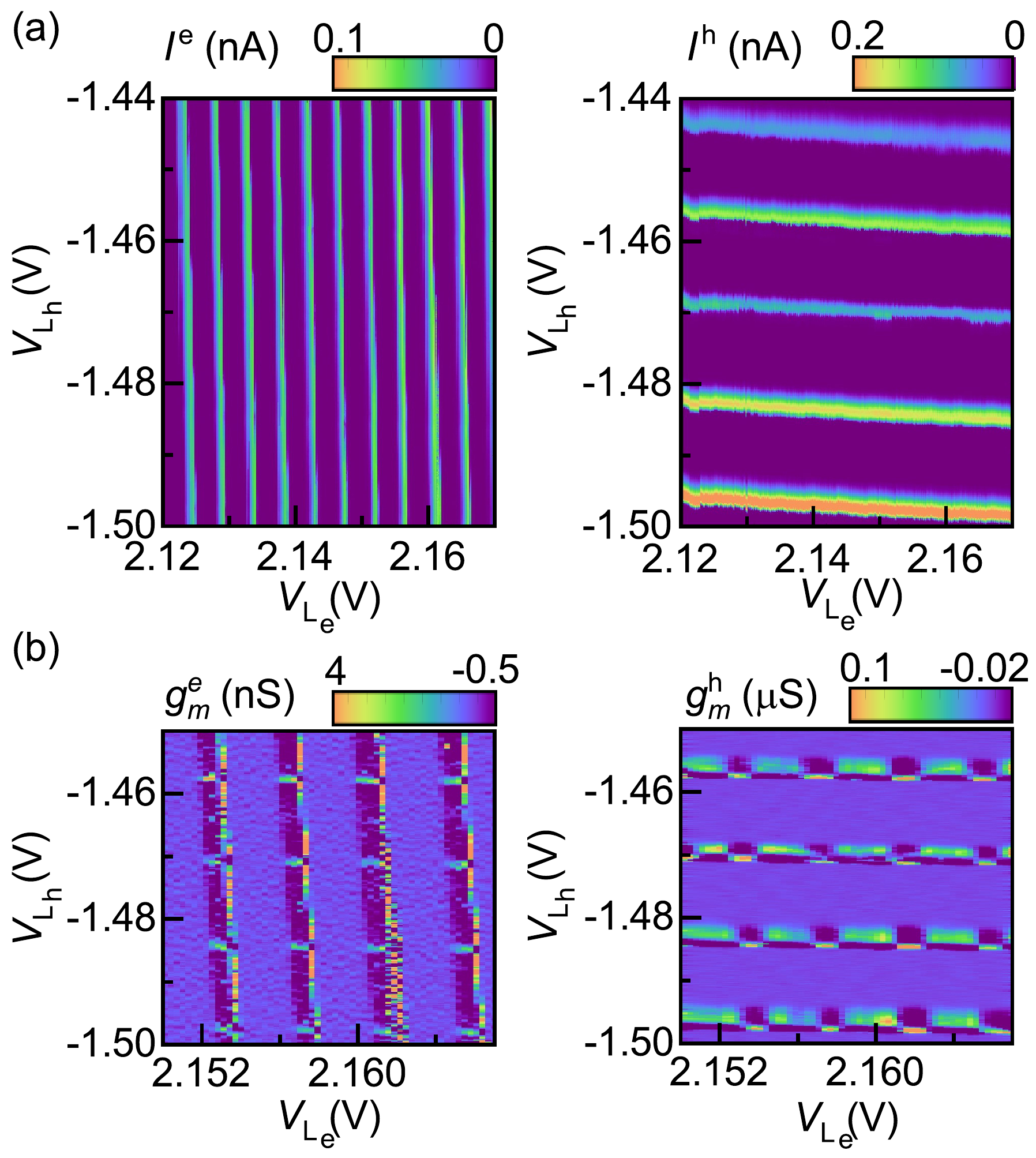}%
\caption{Simultaneous ambipolar charge sensing. (a) Source-drain current versus $V_{\text{L}_e}$ and $V_{\text{L}_h}$ through the SET (right) and the SHT (left). (b) Close-ups of these charge stability diagrams plotted as the tranconductance $g_{\text{m}}^{\text{e}}$ and $g_{\text{m}}^{\text{h}}$ of the SET (right) and of the SHT (left). Data acquired at $V_{\text{B}_{1e}}=1.0$~V, $V_{\text{B}_{2e}}=1.1$~V, $V_{\text{B}_{1h}}=-0.65$~V, $V_{\text{B}_{2h}}=-0.65$~V, and with SET and SHT source-drain voltages fixed at 0.5~mV.  \label{fig:fig3}}
\end{figure}

Previous works have demonstrated simultaneous charge sensing in devices comprising two closely-placed electron QDs~\cite{Podd-APL-2010,Kiyama-SR-2018}. 
In the following we investigate ambipolar simultaneous charge sensing.  
Figure~\ref{fig:fig3}(a) shows charge stability diagrams of the SET and the SHT, measured simultaneously.
The diagrams show 11 electron peaks and 5 hole peaks. If simultaneous charge sensing is taking place, each Coulomb oscillation should exhibit shifts at the intersection of the electron and hole peaks. Such shifts become evident in the plots of the transconductance of the SET ($g_{\text{m}}^{\text{e}}=dI^{\text{e}}/dV_{\text{Lh}}$) and of the SHT ($g_{\text{m}}^{\text{h}}=dI^{\text{h}}/dV_{\text{Le}}$) shown in Fig.~\ref{fig:fig3}(b). Each peak shows a change in sign at the intersections with peaks of the other island. We note that the transconductance measurements shown in Fig.~\ref{fig:fig3}(b) were performed several hours after the acquisition of the charge stability diagrams shown in Fig.~\ref{fig:fig3}(a), thus accounting for the high stability of our device.
The results in Fig.~\ref{fig:fig3} show simultaneous ambipolar charge sensing, i.e. each island is sensing charge transitions in the other.

Having successfully demonstrated charge sensing of the many-electron and many-hole regimes in our ambipolar device, we now aim at detecting few-charge occupation.    
The regular peak spacing in Fig.~\ref{fig:fig3} indicates that both islands are in the many-charge regime. It is fundamental to achieve the few-charge regime in the SET and in the SHT in order to make our ambipolar device suitable for spin manipulation~\cite{Hanson-RMP-2007}. 
The sensitivity of the charge sensing method described above is not uniform but directly proportional to the transconductance of the charge sensor, 
i.e. the slope of the charge sensor peak. Thus, this method of charge sensing is insensitive to transitions when the sensor is in Coulomb blockade.
To overcome this limitation, we implement an active charge sensing method based on the work of Yang~\emph{et al.}~\cite{Yang-AIP-2011}.
Using a computer-controlled dynamic feedback algorithm, we adjust the sensor lead gate $V_{\text{L}}^{\text{S}}$ so that the current through the sensor $I^{\text{S}}$ remains constant at the flank of a Coulomb peak $I^{\text{S}}_0$. On the flank, the transconductance $dI^{\text{S}}/dV_{\text{L}}^{\text{s}}$ and the sensitivity of our sensor are highest. 
Our feedback algorithm takes $I^{\text{S}}$ as the feedback signal and adjusts $V_{\text{L}}^{\text{S}}$ for each data point $x$ measured~\cite{Yang-AIP-2011}

\begin{equation}
\begin{split}
& V_{\text{L}}^{\text{S}}(x+1)=V_{\text{L}}^{\text{S}}(x)-\beta i^{\text{S}}(x)-\Delta V_{\text{s}}A_{\text{C}}(x) \\
& A_{\text{C}}(x+1)=A_{\text{x}}+\frac{\gamma}{\Delta V_{\text{s}}} i^{\text{S}}(x)
\end{split}
\label{eq:feedback}
\end{equation}
where $\Delta V_{\text{s}}=0.1$~mV is the step size of the gate voltage swept as fast axis and $i^{\text{S}}=I^{\text{S}}-I^{\text{S}}_0$ is the error current induced by a charge displacement in the sensed QD. Upon a change in the occupancy of the sensed region, the system experiences a change in the mutual capacitance ratio $A_{\text{C}}=C^{\text{s}}/C_{\text{L}}^{\text{S}}$ between the capacitance of the gate electrode swept as fast axis $C^{\text{s}}$ and the capacitance of the sensor lead gate $C_{\text{L}}^{\text{S}}$.  
The change in $A_{\text{C}}$ is governed by the parameters $\gamma$ and $\beta$, which represent respectively the reactivity of the feedback controller to a sudden change in $i^{\text{S}}$ and the decay rate of the controller to its steady-state. Here, we set $\gamma=0.1~\text{G}\Omega$ and $\beta=2~\text{G}\Omega$

\begin{figure}
\includegraphics[width=15cm]{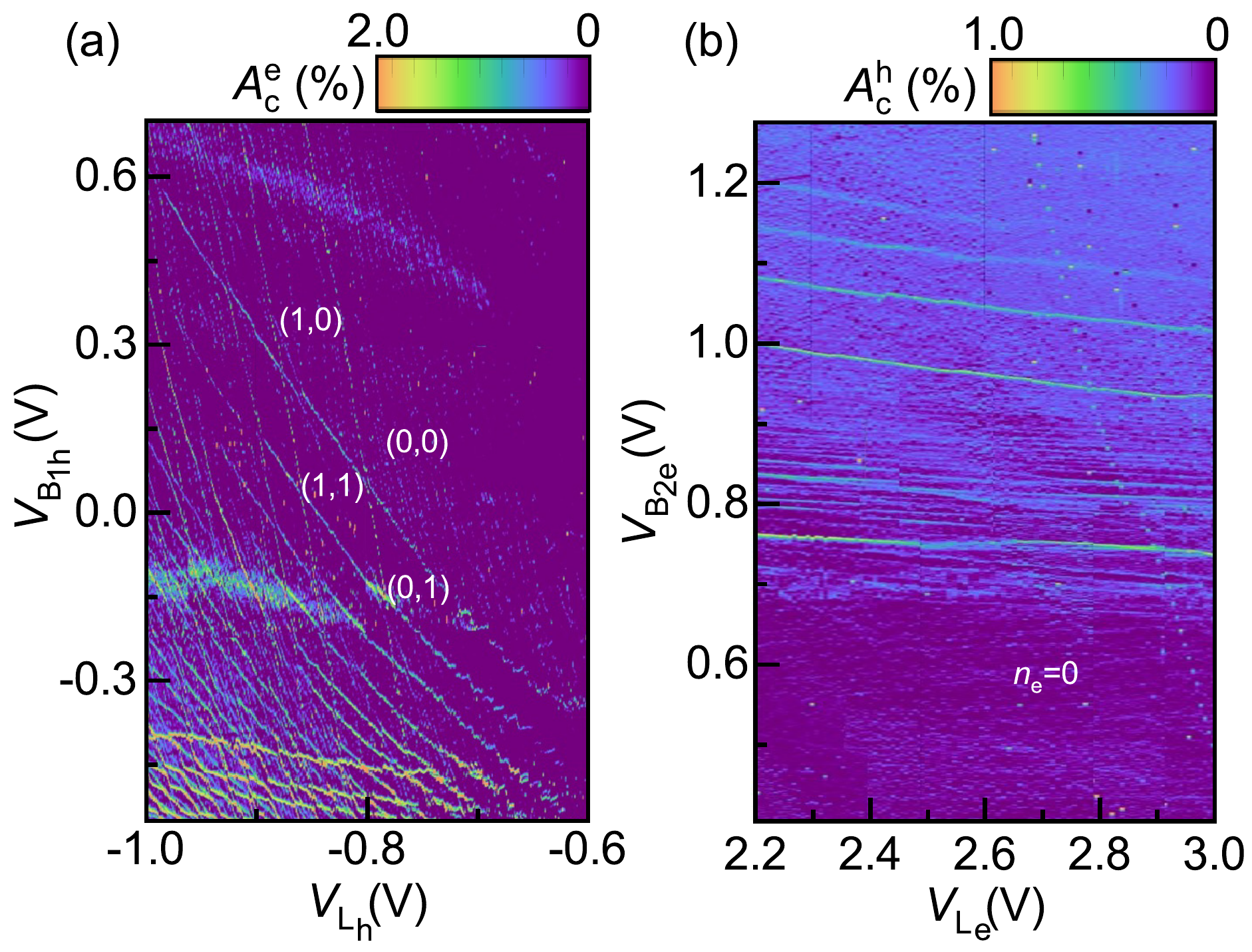}%
\caption{Charge sensing of few-charge occupation in the SET and SHT. Charge stability diagrams of the (a) SHT and (b) SET, extracted by plotting the respective mutual capacitance $A_{\text{C}}$ calculated by the feedback control system. The numbers between brackets indicate the charge occupation of the few-hole double QD. The electron QD is empty at the region indicated by $n_{\text{e}}=0$.  \label{fig:fig4}}
\end{figure}

We use active charge sensing to investigate the few-charge regime of both the SET and the SHT. In Fig.~\ref{fig:fig4} we plot $A_{\text{C}}$ obtained using each of the regions of our device as a charge sensor for the other region. 
Figure~\ref{fig:fig4}(a) shows the charge stability diagram of the SHT using the SET as charge sensor. In this figure, we see a honeycomb pattern of Coulomb oscillations associated to a double quantum dot~\cite{Wiel-RMP-2003}. This double quantum dot is empty at the top right part of the plot ($V_{\text{B}_{1\text{h}}}>0.1$~V and $V_{\text{L}_h}>-0.8$~V) where the SHT region is depleted of holes. 
Figure~\ref{fig:fig4}(b) shows the charge stability diagram of the SET using the SHT as charge sensor. At the bottom-right of this figure ($V_{\text{B}_{2\text{e}}}<0.65$~V) we observe no more electron transitions, which indicates that the SET region is completely depleted of electrons ($n_{\text{e}}=0$). 
We note that Fig.~\ref{fig:fig4} shows additional low-intensity Coulomb oscillations that indicate the presence of unintentional quantum dots in both regions of our device. Such unintentional QDs likely originate from 
disorder caused by defects, e.g. P$_{\text{b}}$-centers~\cite{Spruijtenburg-SR-2016} or chemical alterations of the SiO$_2$ below the gate electrodes~\cite{Brauns-SR-2018}. 
The charge transitions in the plots of $A_{\text{C}}$ are also visible in the plots of $i^{\text{S}}$ with a lower signal-to-noise ratio.
Figure~\ref{fig:fig4} demonstrates few-charge occupation down to the single-charge in both the SET and the SHT of our ambipolar device, achieved using an active charge sensing method. In the SHT region, we identify the formation of a few-hole double quantum dot. For the case of the SET of our ambipolar device, we cannot clearly pinpoint the last electron transition, but for $V_{\text{B}_{2\text{e}}}<0.65$~V the SET is surely depleted.

\section{Conclusion}

In conclusion, we have demonstrated charge sensing between electron and hole quantum dots in silicon and succesfully tuned each QD to the few-charge regime. To date, ambipolar devices had only been studied via direct transport measurements. To enable the sensitivity needed to detect the few-charge regimes in our device, we implemented active charge sensing. 
This technique can be further used for spin readout, as reported for electron~\cite{Morello-Nat-2010} and hole~\cite{Crippa-NatC-2019} spin qubits.
Detecting the few-charge regime is a crucial step towards ambipolar spin qubits in silicon.
Such devices provide a means to combine the readout possibilities of a 
single-electron transistor charge sensor~\cite{Morello-Nat-2010} with the favorable qubit properties of holes, namely strong spin-orbit coupling for all-electrical spin manipulation~\cite{Nowak-Sci-2007} and suppressed hyperfine interaction with nuclear spins of the host material~\cite{Bulaev-PRB-2005}. 
Ambipolar spin qubits may also be interconnected in linear arrays of multiple tunnel-coupled electron and hole spin qubits which can be reconfigured on-the-fly in a way similar to standard CMOS circuits.
Another promising use for ambipolar devices can be as converters of spin to spin-polarized light. Our present device architecture could enable spin-dependent electron and hole recombination via charge transfer between the two neighbouring QDs. While exciton recombination in silicon is hindered by its indirect band gap, this might be mitigated in nanosize silicon, as studies have reported that the band gap of silicon can be made direct via quantum confinement~\cite{Dohnalova-Light-2013} or via growth in a hexagonal crystalline structure~\cite{Guo-SR-2015,Hauge-NL-2015}.

\end{document}